\documentclass[10pt,letterpaper]{article}
\usepackage{opex3}

\usepackage{cite}
\usepackage{subfigure}

\begin{document}

\title{Ultrafast quantum random number generation based on quantum phase fluctuations}

\author{Feihu Xu,$^{1, *}$ Bing Qi,$^{1}$ Xiongfeng Ma,$^{1, 2}$ He Xu,$^{1}$ Haoxuan Zheng,$^{1, 3}$ and Hoi-Kwong Lo$^{1}$}

\address{%
$^{1}$ Center for Quantum Information and Quantum Control, \\ Department of Physics and Department of Electrical \& Computer Engineering, \\ University of Toronto, Toronto, ON, M5S 3G4, Canada \\
$^{2}$ Present address: Center for Quantum Information, Institute for Interdisciplinary Information Sciences, Tsinghua University, Beijing, China \\
$^{3}$ Present address: Department of Physics and MIT Kavli Institute, Massachusetts Institute of Technology, 77 Massachusetts Ave., Cambridge, MA 02139, USA}

\email{$^{*}$feihu.xu@utoronto.ca}

\date{\today}

\begin{abstract}
A quantum random number generator (QRNG) can generate true
randomness by exploiting the fundamental indeterminism of quantum
mechanics. Most approaches to QRNG employ single-photon detection
technologies and are limited in speed. Here, we experimentally
demonstrate an ultrafast QRNG at a rate over 6 Gbits/s based on the
quantum phase fluctuations of a laser operating near threshold.
Moreover, we consider a potential adversary who has partial
knowledge on the raw data and discuss how one can rigorously remove
such partial knowledge with post-processing. We quantify the quantum
randomness through min-entropy by modeling our system and employ two randomness
extractors - Trevisan's extractor and Toeplitz-hashing - to distill
the randomness, which is information-theoretically provable. The
simplicity and high-speed of our experimental setup show the
feasibility of a robust, low-cost, high-speed QRNG.
\end{abstract}



\section{Introduction}
Random numbers play a key role in many areas, such as statistical
analysis, computer simulations and cryptography
\cite{meteopolis1949monte, bennett1984quantum, schneier1995applied}.
Traditionally, pseudo-random number generator (pseudo-RNG) based on
deterministic algorithms has long been used for various
applications. Recently, physical-RNG based on chaotic behaviors of
semiconductor lasers has been proposed to generate ultrahigh-speed random bits
\cite{uchida2008fast, reidler2009ultrahigh, kanter2010}. Generally speaking,
the above schemes cannot generate truly random numbers with information-theoretically provable randomness. The output signal from a chaotic-laser system \cite{uchida2008fast, reidler2009ultrahigh, kanter2010} presents a property of periodicity due to the photon round trip time and the chaotic-laser system is essentially a deterministic system, which in principle cannot offer truly inherent randomness. We remark that very recently, fast physical-RNG using amplified optical noise has also been demonstrated \cite{Williams2010, li2011scalable}.

Quantum-RNG (QRNG), on the other hand, can generate truly random numbers
from the fundamentally probabilistic nature of quantum processes. In
the past decade, several QRNG schemes, such as single-photon
detection \cite{jennewein2000fast, dynes2008high, Kwiat2010, Furst2010, Wahl2011}, quantum
non-locality \cite{pironio2010random}, and vacuum state fluctuations
\cite{gabriel2010generator} have been demonstrated. Meanwhile,
commercial QRNGs such as ID Quantique system \cite{IDQ} have already
appeared on the market. However, previous implementations have been
limited to a relatively low rate due to the difficulties in measuring quantum effects: the speed of the single-photon-detection QRNG \cite{jennewein2000fast, dynes2008high, Kwiat2010, Furst2010, Wahl2011} is limited by the maximum counting rate of single-photon detectors, which is typically below 100 MHz \cite{Hadfield2009}; quantum-non-locality QRNG \cite{pironio2010random} is a proof-of-concept demonstration (with a random number generation rate on the order of 1 bit/s) and thus unsuitable for practical applications; building a fast shot-noise limited homodyne-detector for vacuum-state-fluctuations QRNG \cite{gabriel2010generator} is also a big challenge. In 2009, our group proposed and built a fast QRNG by measuring the quantum phase fluctuations (or noise) of a laser, which yields a speed of 500 Mb/s \cite{QiAQIS, Qi2010}. Instead of directly measuring weak quantum effects, this scheme measures the enhanced quantum noise (amplified spontaneous emissions) and thus can be realized by \emph{conventional photodetectors} at a high-speed and with a low cost. A similar scheme at a lower speed has also been demonstrated \cite{guo2010truly}. Nonetheless, the key point is, the generation rates of all previous QRNGs are still too low for many applications, such as high-speed quantum key distribution \cite{takesue2007quantum} operating over GHz.

Moreover in real experiments, the quantum randomness is inevitably
mixed with the classical noise, which may be observed or even
controlled by a potential adversary. If we consider a scenario where
the adversary tries to guess the outcomes from a QRNG, then she
could take advantage of the side information due to classical noise.
Thus, a refined post-processing scheme is necessary to remove the correlation between the generated random bits and the classical noise. Two post-processing methods that are widely-used in various QRNG implementations are least-significant-bits (LSB) \cite{uchida2008fast, reidler2009ultrahigh, kanter2010, guo2010truly} and non-universal hashing functions \cite{gabriel2010generator, Kwiat2010, Wahl2011}. It is important to perform such post-processing on the raw data to distill out a shorter, but more secure, string of random bits. However, neither LSB procedures nor non-universal hashing functions are \emph{information-theoretically provable}, the property of which is especially valuable in current technology. Hence, it is still arguable whether these two methods can indeed extract out perfect-random bits. On the other hand, in theoretical computer
science, there has been much interest in post-processing methods,
called randomness extractors \cite{WC_Authen_81, Trevisan:Extractor:1999, shaltiel2004recent}. The randomness from many extractors has been information-theoretically proven, such as Trevisan's extractor \cite{Trevisan:Extractor:1999}. However, none of these extractors
have been implemented in a real QRNG experiment. Therefore, there is
a gap between theory and experiment.

In this paper, we report an ultrafast QRNG with a generation rate
over 6 Gb/s based on measuring the quantum phase fluctuations of a
laser operating at a low intensity level. Compared with our
previous works \cite{QiAQIS, Qi2010}, both the hardware design and
post-processing algorithm have been substantially improved. On the
hardware side, a compact planar lightwave circuit Mach-Zehnder
interferometer (PLC-MZI) with internal temperature control is
introduced to replace the bulky MZI constructed with discrete fiber
components in Ref.~\cite{QiAQIS, Qi2010}. The high stability of the
PLC-MZI allows us to stabilize its phase by simply controlling its
temperature. We emphasize that the simple and robust design of our
QRNG suggests that it can be readily commercialized for practical
applications.

On the post-processing side, we bridge the gap between the theory
and practice of randomness extraction by applying randomness
extractors. Here, we quantify the quantum randomness and classical
noise (present in the QRNG) separately by min-entropy. The
min-entropy is defined as $H_\infty(X) = - \log_2\left(\max_{x\in\{0,1\}^n}Pr[X=x] \right)$, which quantifies the amount of randomness of a distribution $X$ on $\{0,1\}^n$. In our experiment, the min-entropy is evaluated by modeling the physical setup.
We optimize the experimental parameters by maximizing the quantum randomness and implement two randomness extractors, Trevisan's extractor
\cite{Trevisan:Extractor:1999} and Universal hashing
(Toeplitz-hashing) \cite{WC_Authen_81}. Both methods take
finite-size effects into consideration. With this new
post-processing scheme, we not only improve the random number
generation rate \cite{QiAQIS, Qi2010} by more than one order of magnitude, but
also achieve an information-theoretically provable randomness. It is
the first QRNG experiment that implements such extractors.

\section{Experimental demonstration}
%
It is well known that the fundamental phase fluctuations (or noise)
of a laser can be attributed to spontaneous emission, which is
quantum mechanical by nature \cite{henry1982theory}. The quantum
phase fluctuations are inversely proportional to the laser output
power \cite{henry1982theory}. By operating the laser at a low
intensity level, the quantum phase fluctuations can dominate over
classical phase noise \cite{Yariv1983} and be readily extracted to
generate truly random numbers.

\begin{figure}[tbh]
\centering \resizebox{8.5cm}{!}{\includegraphics{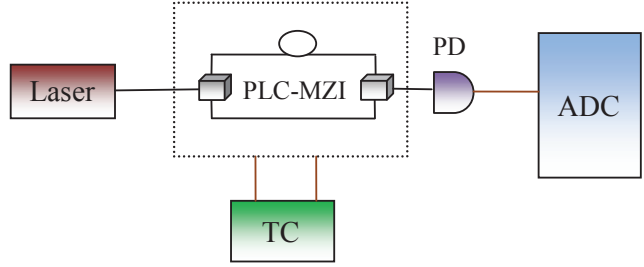}}
\caption{Experimental setup. Laser, 1550nm cw  DFB laser diode (ILX
Lightwave); PLC-MZI, planar lightwave circuit Mach-Zehnder
interferometer with a 500ps delay difference (manufactured by NTT);
TC, temperature controller (PTC 5K from Wavelength Electronics
Inc.); PD, 5GHz InGaAs photodetector (Thorlabs SIR5-FC); ADC, 8-bit
analog-to-digital convertor inside an oscilloscope (Agilent
DSO81204A).} \label{Fig.1}
\end{figure}

We have developed a delayed self-heterodyning system to measure the
phase fluctuations. The schematic diagram of the experimental setup
is shown in Fig.~\ref{Fig.1}. A 1.55 \emph{$\mu$m} single mode cw distribute-feedback (DFB) diode laser (ILX lightwave) operating at a low intensity level
is the source of quantum phase fluctuations. A PLC-MZI with a 500ps
delay difference (manufactured by NTT) is employed to convert the
phase fluctuations to intensity fluctuations, which is subsequently
detected by a 5GHz InGaAs photodetector (Thorlabs). Note that to
achieve a high interference visibility, a polarization maintaining
fiber is used to connect the laser and the PLC-MZI. A temperature
controller (TC) \cite{temperature:controller} is used to stabilize
the phase difference of PLC-MZI. The photodetector output is further sampled and
digitized by an 8-bit analog-to-digital convertor (ADC) to generate
random bits.

The electric field of the DFB laser beam is given by
\begin{equation} \label{electricfield}
E(t)=E_{0}\exp[i(\omega t+\theta(t))]
\end{equation}
where $\theta(t)$ represents the phase fluctuations of the laser source and
$\omega$ is angular frequency. By stabilizing the phase difference of PLC-MZI (the phase delay introduced by the path length difference of the two arms) at $[2m\pi+\pi/2]$
(where $m$ is an integer), the output voltage $V(t)$ from the
photodetector (after removing a DC background) can be described by
\cite{henry1982theory, petermann1988laser}
\begin{equation} \label{QRNG:Model:beatsignal}
V(t) \propto 2E(t)E(t+\tau)\sin(\Delta\theta(t)) \propto
P\Delta\theta(t)
\end{equation}
where $P$ is the laser output (emission) power, $\tau=500ps$ is the constant time delay between the two arms of our PLC-MZI, and $\Delta\theta(t)=\theta(t)-\theta(t+\tau)$ is the total phase fluctuations measured by the interferometric system. Strictly speaking, $\Delta\theta$ is a function of both $t$ and $\tau$. Since $\tau$ is a constant in our system, we treat $\Delta\theta$ as a function of $t$ for simplicity. Here, $\Delta\theta(t)$ is sufficiently small such that $\sin(\Delta\theta(t))\approx \Delta\theta(t)$. We have assumed that the intensity noise of the laser is negligible \cite{henry1982theory}, which has been verified experimentally (see discussion below).

It is convenient to further separate the total phase fluctuations (measured by the system) into a quantum part and a classical part. While the quantum phase
fluctuations are inversely proportional to laser output power and
can be treated as Gaussian white noise \cite{petermann1988laser},
the classical phase noise is laser-power independent
\cite{classicalnoise}, which in principle could be controlled by
a potential adversary. Hence, the variance of the total phase fluctuations can be written as
\begin{equation} \label{phasefluctuations}
\langle\Delta\theta(t)^{2}\rangle=\frac{Q}{P}+C
\end{equation}
where $\langle\bullet\rangle$ denotes a statistical average, $\frac{Q}{P}$ and $C$ represent the contributions of quantum phase fluctuations and classical phase noise respectively. We remark that within the time scale of our experiment, $Q$, $P$ and $C$ do not vary with time.

In practice, the detection system will also contribute a laser-power
independent background noise $F$. Therefore, the variance of the output a.c. voltage $V_{pr}(t)$ from the photodetector of our system is given by
\begin{equation}  \label{voltvariance}
\langle V_{pr}(t)^{2}\rangle=AQP+ACP^{2}+F
\end{equation}
where $A$ is is a constant determined by the gain of the photodetector.

In Eq.~(\ref{voltvariance}), the term $AQP$ is quantum fluctuations
part, from which true randomness can be extracted. We name it \emph{quantum signal}. On the other hand, the term $ACP^{2}+F$
quantifies \emph{classical noise} due to technical imperfections
that potentially could be controlled by an adversary. In
principle, the amount of extractable quantum randomness is
independent of the magnitude of classical noise. However in
practice, it is challenging to extract a small quantum signal on top
of a large classical noise background. To generate high-quality
random numbers, we would like to maximize the quantum signal while
keeping the classical noise as low as possible.

One common figure of merit in signal processing is the
signal-to-noise ratio (SNR). In our QRNG system, SNR $\gamma$ can be defined as
\begin{equation}  \label{SNR:eqn}
\gamma=\frac{AQP}{ACP^{2}+F}
\end{equation}
Given parameters $AQ$, $AC$, and $F$, we can choose a suitable laser power $P$ to maximize $\gamma$.

\begin{figure}[!t]
\centering \resizebox{10cm}{!}{\includegraphics{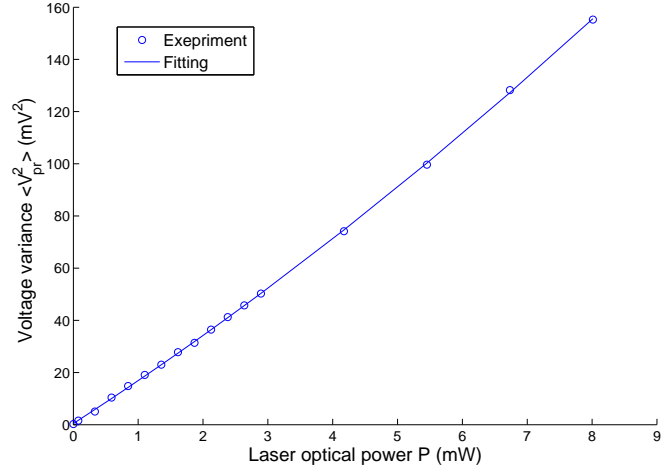}}
\caption{Experimental voltage variance. The variance of the output a.c. voltage ($V_{pr}(t)$) from the photodetector (see Fig.~\ref{Fig.1}) is measured by an oscilloscope. Here, the error bars are smaller than the symbol size. The experimental data is fitted by a quadratic polynomial function. } \label{fitting}
\end{figure}

To determine the parameters $AQ$, $AC$, and $F$ experimentally, we measured the variance of $V_{pr}(t)$ at different optical
power levels (see Fig.~\ref{fitting}) and then fit the experimental data (with least square
estimation fitting) using Eq.~(\ref{voltvariance}). The experimental
results and the corresponding confidence intervals (level
$\alpha=0.99$) are shown in Table \ref{tab2}.

\begin{table}[!h]\center
\caption{Experimental results (with $0.99$ confidence intervals) of
parameters in Eq.~(\ref{voltvariance}).}\label{tab2}
\begin{tabular}{c @{\hspace{0.8cm}} c @{\hspace{0.8cm}} c}
\hline F ($mV^{2}$) & AQ ($mV^{2}/mW$) & AC ($mV^{2}/mW^{2}$) \\
       $0.36\pm0.06$ & $16.1\pm0.5$ & $0.4\pm0.2$ \\
\hline
\end{tabular}
\end{table}

\begin{figure}[!t]
\centering \resizebox{8cm}{!}{\includegraphics{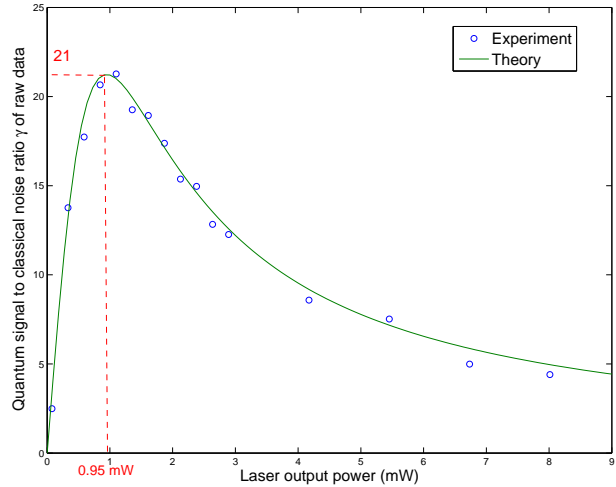}}
\caption{Quantum signal to classical noise ratio. The theoretical
curve of signal-to-noise ratio is obtained from Egn.~\ref{SNR:eqn} and the parameters given in Table~\ref{tab2}. The experimental
results are measured with an oscilloscope at different laser powers \cite{experiment:gamma} (the corresponding error bars are smaller than the symbol size). At low and high power range, either the background noise $F$ or the classical phase noise $ACP^{2}$ will dominate over the
quantum signal. The optimal ratio $\gamma=21$ is achieved at
$P=0.95$ $mW$.} \label{ratio}
\end{figure}

Using Eq.~(\ref{SNR:eqn}) and the data given in Table~\ref{tab2}, we calculate $\gamma$ as a function of laser power. The results are shown in
Fig.~\ref{ratio}. The experimental data points are determined with an oscilloscope at different laser
power levels (see details in Ref.~\cite{experiment:gamma}). We can see that at low and high power range, either the background
noise $F$ or the classical phase noise $ACP^{2}$ will dominate over
the quantum signal. The optimal ratio $\gamma=21$ is achieved at
$P=0.95$ $mW$. As shown in Ref.~\cite{Working:QRNGpp:2011}, by
operating the laser at this power, the extractable quantum
randomness is also maximized. Therefore, we choose $0.95$ $mW$ as
the working point of laser.

\begin{figure}[!t]
\centering
\resizebox{8cm}{!}{\includegraphics{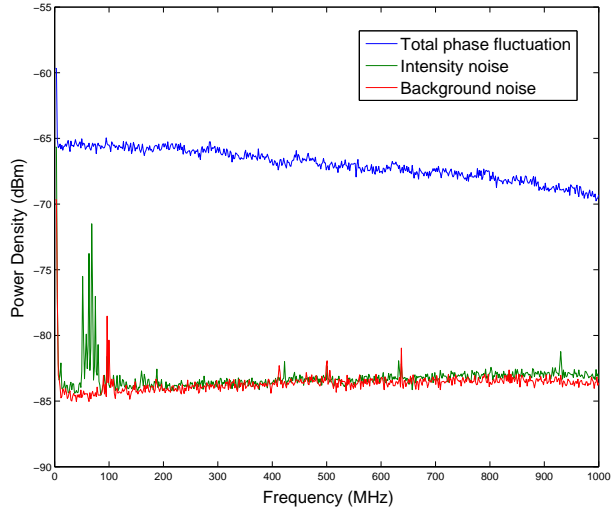}} \caption{Noise spectra. The spectral power density of total
phase fluctuations (blue), intensity noise (green), and background
noise (red).} \label{spectrum}
\end{figure}

We also perform measurements in the frequency domain by using an RF
spectrum analyzer. Three different power-spectra have been acquired:
(1) the spectrum of total phase fluctuations under the optimal working
condition ($0.95$ $mW$); (2) the intensity noise spectrum acquired
by connecting the laser output directly to the photodetector; (3)
the background noise spectrum acquired by turning off the laser. The results are shown in Fig.~\ref{spectrum}. We can see
that under the normal operating condition, the intensity noise is
negligible compared to the phase fluctuations. This result supports
our previous assumption. Note that since the PLC-MZI is a fiber-pigtailed compact device, we expect that the coupling efficiency between the laser and the PLC-MZI will stay constant over time. As we expect from a perfect white noise source, the spectrum of phase fluctuations itself is flat over the
whole measurement frequency range. In the spectrum of background noise, there are a few spectral lines that could be environmental EM
noise picked up by our detector \cite{spectrum:peaks}.

%
The experimental procedure for random number generation is as
follows. The laser output power is set to $0.95$ $mW$ by adjusting
its driving current. The TC is carefully adjusted to stabilize the
phase difference of PLC-MZI at $[2m\pi+\pi/2]$. The output a.c. voltage ($V_{pr}(t)$) from the photodetector is sampled and digitized at 1 GHz sampling rate \cite{sampling:rate} with an 8-bit ADC. The sampling range of ADC is determined by the peak-to-peak voltage of $V_{pr}(t)$ (around 30 mV) \cite{adc:range} and the bins of ADC are equally spaced. Thus, from each sample point, we can generate 8 raw random bits that are ready for post-processing.

\section{Min-entropy evaluation} \label{minentropy}
As mentioned earlier, the raw random bits from our QRNG are contributed by both the
quantum signal and the classical noise. In order to remove the correlation between the random bits and the classical noise (and thus extract pure quantum randomness), we apply a post-processing scheme that is composed of two main parts - quantum min-entropy (or randomness) evaluation and
randomness extraction. The evaluation of min-entropy is discussed in this section, while randomness extraction will be described in the following section.

\begin{figure}[!t]
\centering \resizebox{9cm}{!}{\includegraphics{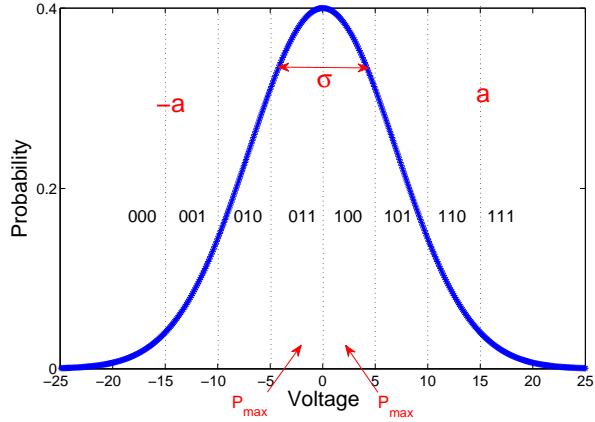}}
\caption{A simple illustration of min-entropy evaluation (toy model). The raw-data follows a Gaussian distribution ($\mu=0$ and $\sigma=7$) and is digitized by a 3-bit ADC (sampling range is defined as [-a, a] with a=15 here). From Eq.~(\ref{minentropy:definition}), the min-entropy is determined by the sample point $x$ with the maximal probability $P_{max}$. Here, $x$ equals to the bin of `100' (or `011') and $P_{max}$ can be calculated from its bin area.} \label{minentropy:fig}
\end{figure}

Min-entropy is defined as
\begin{equation} \label{minentropy:definition}
H_\infty(X) = - \log_2\left(\max_{x\in\{0,1\}^n}Pr[X=x] \right)
\end{equation}
It quantifies the amount of randomness of a distribution $X$ on $\{0,1\}^n$. From Eq.~(\ref{minentropy:definition}), the min-entropy of a given sequence $X$ is determined by the sample point $x$ with maximal probability $P_{max}=\max_{x\in\{0,1\}^n}Pr[X=x]$. A simple illustration of the evaluation process is shown in Fig.~\ref{minentropy:fig}, where the raw-data follows a Gaussian distribution and is digitized by a 3-bit ADC. Hence, the raw-data will be mapped to a binary sequence $X$ with 3 dimensions ($n$=3 in Eq.~(\ref{minentropy:definition})). The sample point (one of the 8 bins in Fig.~\ref{minentropy:fig}) with maximal probability is `011' (or `100') and its corresponding probability $P_{max}$ can be calculated from its bin area. Note that in Fig.~\ref{minentropy:fig}, three key parameters to determine $P_{max}$ (thus min-entropy) are the standard deviation of Gaussian distribution ($\sigma$), the ADC sampling range ($a$) and the resolution of ADC (3-bit in Fig~\ref{minentropy:fig}).

In our experiment, a physical model is derived to evaluate the quantum min-entropy of the raw-data. Our main assumptions are: (1) Quantum signal is independent of classical
noise when the laser is operating above threshold; (2) Quantum signal follows a Gaussian distribution \cite{petermann1988laser}; (3) The sequence of the raw-data is independent and identically distributed (iid, See \cite{sampling:rate} and discussion below).

With these assumptions, we can calculate the quantum min-entropy of the raw-data as follows.
\begin{enumerate}
\item
Determine the sampling range and evaluate the total variance: the
working range of sampling system (8-bit ADC in Fig.~\ref{Fig.1}) $a$ is
determined by the output voltage from the photodetector ($V_{pr}(t)$ in Eq.~(\ref{voltvariance})). From random sampling, we can obtain the variance of the total fluctuations, $\sigma^{2}_{total}=AQP+ACP^{2}+F$. At the laser emission power 0.95 $mW$, we choose the ADC sampling range as $a=15$ mV and obtain the variance of the total fluctuations as $\sigma^{2}_{total}=24.4 mV^{2}$.

\item
Evaluate signal to noise ratio: from Fig.~\ref{ratio}, we evaluate
the quantum signal to classical noise ratio. At 0.95 $mW$, the ratio
is $\gamma=21$.

\item
Evaluate the quantum variance: from Eq.~(\ref{SNR:eqn}) and step 1 and 2, we can calculate
the variance of the quantum signal (following a Gaussian distribution) as $AQP$.  At 0.95 $mW$, the
quantum variance is $\sigma^{2}_{quantum}=\frac{\gamma}{\gamma+1}\sigma^{2}_{total}=23.3 mV^{2}$.


\item
Calculate the lower bound of the quantum min-entropy: as shown in Fig.~\ref{minentropy:fig}, given the ADC range $a$, we evaluate the bin with maximal probability from the Gaussian distribution derived in Step 3, which gives the lower bound of the min-entropy of the quantum signal. Note that in real experiment we use an 8-bit ADC (instead of 3-bit in Fig.~\ref{minentropy:fig}) and its bins are equally spaced. At 0.95 $mW$, the standard deviation of the Gaussian distribution is $\sigma_{quantum}=4.8$ $mV$ and the corresponding maximal probability of the raw sequence is $P_{max}=9.6\times 10^{-3}$. Therefore, from Eq.~(\ref{minentropy:definition}), the quantum min-entropy of our raw-data is 6.7 bits per sample (8 raw bits from the ADC in Fig.~\ref{Fig.1}).
\end{enumerate}

One might ask `whether the randomness generation rate is bounded by
the channel capacity \cite{channelcapacity} of a noisy channel
through the signal-to-noise ratio?' In our opinion, the answer is
\emph{no}. The main function of a QRNG is not to recover the quantum signal
(source of quantum randomness) from the background of the classical
noise, but to generate random bits which have no correlations with
the classical noise. We can assume that our quantum signal
ultimately gives us a classical number \emph{X} (random number generation) and the adversary (Eve) inputs a classical number \emph{E} (by controlling
the classical noise). \emph{X} and \emph{E} are combined together to obtain the
raw output \emph{Y}. Hence, the whole discussion can be phrased within
classical information theory. For instance, we can consider the
discrete case and the simple function that \emph{Y} is the exclusive-or of
\emph{X} and \emph{E}. Suppose \emph{X} and \emph{E} are both one random bit, then the output \emph{Y}
will also be random and the conditional entropy $H (Y|E)$ is 1 bit,
where Eve has no information about \emph{Y}. On the other hand, since Eve
can decide whether to flip the bit or not, the mutual information
between \emph{X} and \emph{Y} is $I(X:Y)=0$. The channel capacity is zero. In
summary, this simple example shows that it is possible for the
randomness generation (i.e. $X=1$ bit) to be non-zero even though the
channel capacity is zero. The channel capacity is not an upper bound
to the randomness generation rate.

\section{Randomness extraction and statistical tests}
In previous section, we have shown that the lower bound of quantum min-entropy is 6.7 bits per sample, which means that we can generate 6.7 information-theoretically random bits from each sample (8 raw bits). We remark that the raw-data itself cannot pass the statistical random tests, which is mainly due to the classical noises mixed in the raw-data and the fact that the as-obtained quantum phase fluctuations follow a Gaussian distribution instead of an uniform distribution. Therefore, to extract the 6.7 perfect-random bits and improve the randomness quality of our raw-data, randomness extractor is implemented. Roughly speaking, a randomness extractor is a function as
\begin{equation} \label{Ext:Def:extractor}
\{0,1\}^n\times\{0,1\}^d\rightarrow\{0,1\}^m
\end{equation}
which means that for a raw and non-perfect-random sequence $X$ on $\{0,1\}^n$ with min-entropy $H_\infty(X)\ge m$, the extracted output sequence $Y$ is a nearly uniform distribution on $\{0,1\}^m$. In other words, a randomness extractor takes a small random seed ($d$ bits) and a raw random source ($n$ bits) and outputs a near perfect-random bit-string ($m$ bits). A more rigorous discussion of randomness extractor can be found in \cite{Working:QRNGpp:2011}.

We implement two randomness extractors, Toeplitz-hashing extractor \cite{WC_Authen_81} and Trevisan's extractor \cite{Trevisan:Extractor:1999}. Both are proven to be information-theoretically secure and take finite-size effects into account \cite{finitesize}. The details of our implementation are discussed in Ref.~\cite{Working:QRNGpp:2011}. Here, we give a brief summary. Toeplitz-hashing extractor extracts random bit-string $m$ by multiplying the raw sequence $n$ with the Toeplitz matrix ($n$-by-$m$ matrix, random seed). The seed length of random bits required to construct the Toeplitz matrix is $d=n+m-1$ \cite{Krawczyk1994}. In our implementation with Matlab on a standard laptop computer, we choose the input bit-string length as $n=4096$. Since the min-entropy of our raw-data is 6.7 bits per 8-bit sample, the output bit-string is $4096\times6.7/8\ge3430$. To generate near perfect-random bits, the output length is set to $m=3230$ bits (see \cite{Working:QRNGpp:2011}). Hence, a $4096$-by-$3230$ Toeplitz matrix is generated \cite{seed:generate} in construction of the Toeplitz-hashing extractor, which achieves a speed of 441 Kb/s. Note that Toeplitz-hashing can be implemented much faster with hardware implementations \cite{Krawczyk1994}. For Trevisan's extractor, we implement its improved version \cite{Raz1999} and the details are shown in \cite{Working:QRNGpp:2011}.


%

\begin{figure*}[htb]
\centering \subfigure[] {\includegraphics [width=5.5cm]
{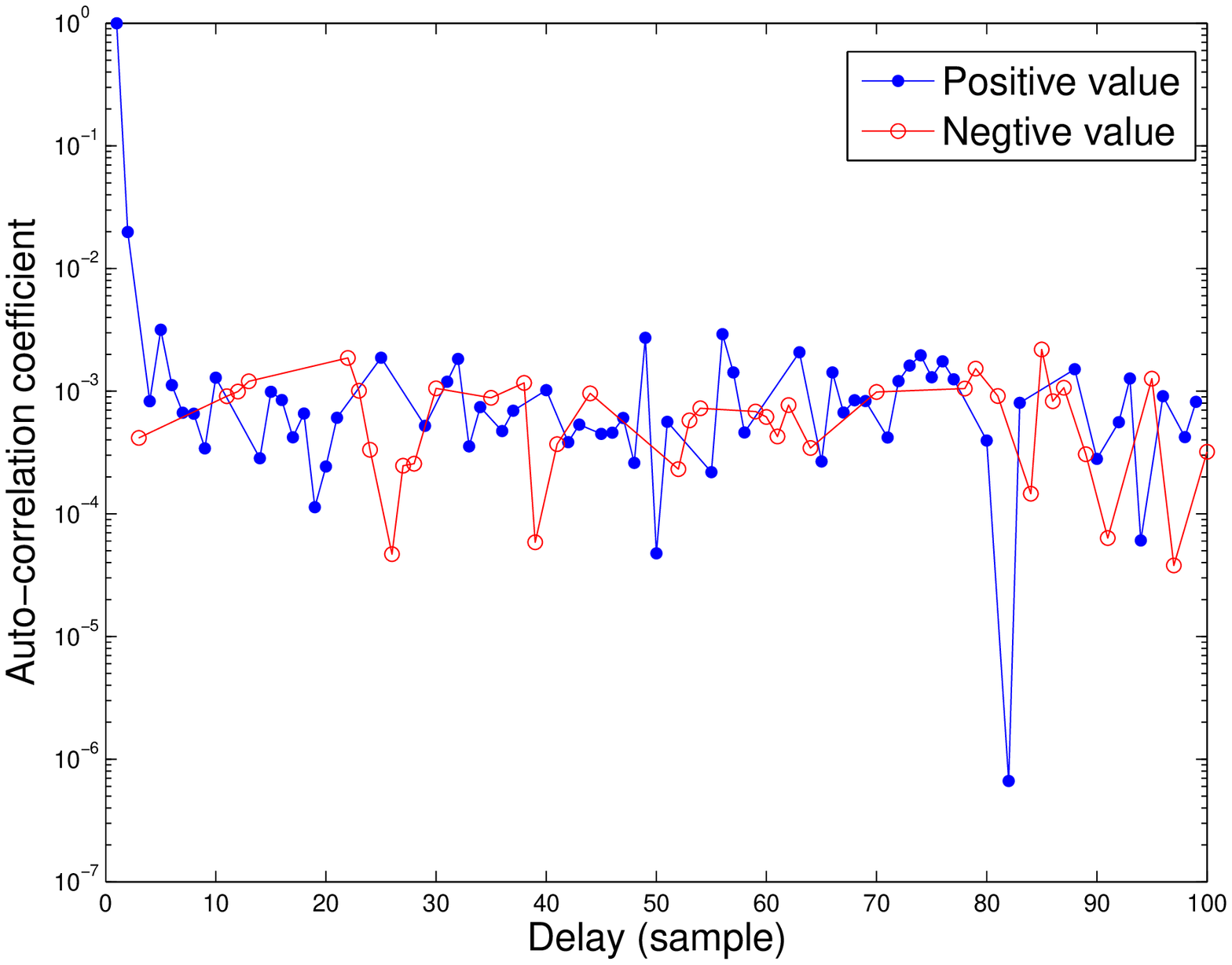} \label{fig:subfig1}} \qquad \subfigure[]
{\includegraphics [width=5.5cm] {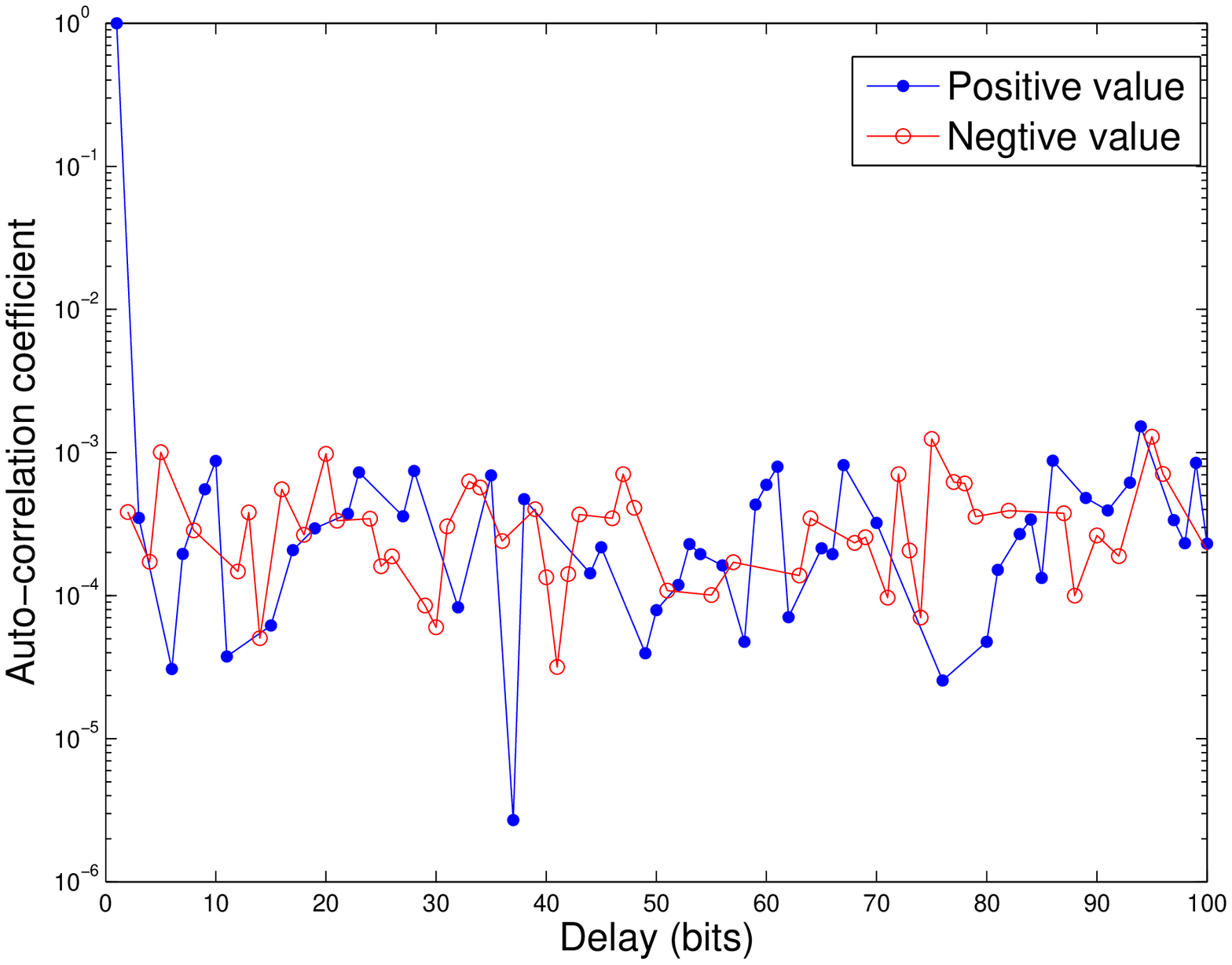}
\label{fig:subfig2}} \qquad \subfigure[] {\includegraphics
[width=5.5cm] {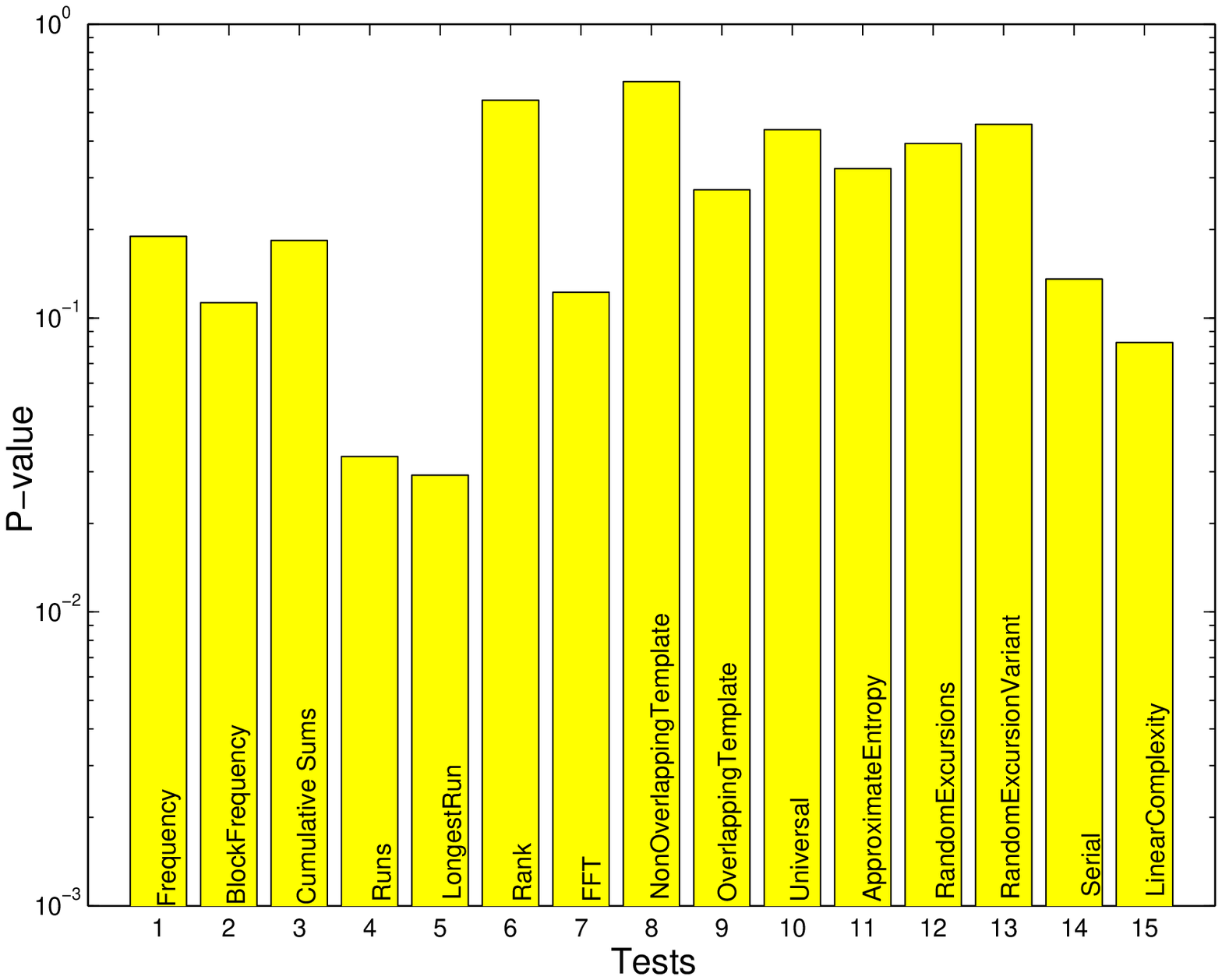} \label{fig:subfig3}} \qquad \subfigure[]
{\includegraphics [width=5.5cm] {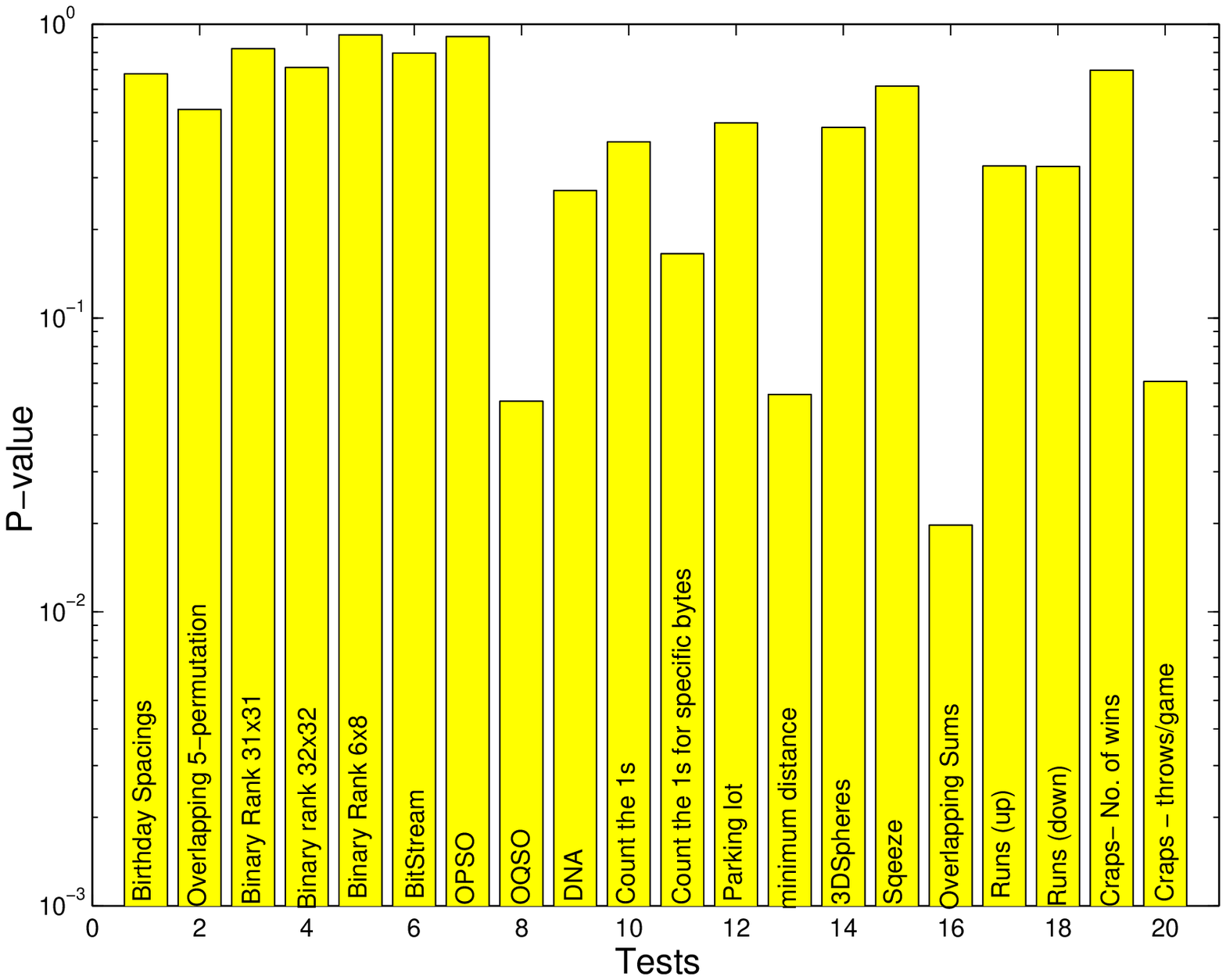} \label{fig:subfig4}}
\caption{ (a) Autocorrelation of the raw-data. The raw-data is obtained by sampling the output a.c. voltage ($V_{pr}(t)$ from the photodetector) with an ADC (see Fig.~\ref{Fig.1}). Each sample consists of 8 bits and the correlation between samples cannot reach zero for a practical detector with finite bandwidth. (b) Autocorrelation of the Toeplitz-hashing output. Data size is $1\times10^{7}$ bits and the average value within 100 bit-delay is $-1.0\times10^{-5}$. In theory, for a truly random $1\times10^{7}$ bit string, the average normalized correlation is 0 and the standard deviation is $3\times10^{-4}$. In practice, due to the inevitable presence of bias and finite data size, the autocorrelation of data sequence can never reach 0. (c) NIST results of the Toeplitz-hashing output. Data size is 3.25 Gbits (500 sequences with each sequence around 6.5 Mbits). To pass the test, P-value should be larger than the lowest significant level $\alpha=0.01$, and the proportion of sequences satisfying $P > \alpha$ should be greater than 0.976. Where the test has multiple P-values, the worst case is selected. (d) Diehard results of the Trevisan's extractor output. Data size is 240Mbits. A Kolmogorov-Smirnov (KS) test is used to obtain a final P-value from the case of multiple P-values. Successful P-value is $0.01\leq P\leq0.99$.}
\end{figure*}

The output from both extractors successfully passes all the standard statistic tests of Diehard \cite{Diehard}, NIST \cite{NIST}, and TestU01 \cite{testu01}. The autocorrelations of the raw-data and the Toeplitz-hashing output are shown in Fig.~\ref{fig:subfig1} and Fig.~\ref{fig:subfig2}. Here, the autocorrelation coefficient $R$ of a sequence $X$ is defined as
\begin{equation} \label{Ext:autocorrelation}
R(j)= \frac{E[(X_{i}-\mu)(X_{i+j}-\mu)]}{\sigma^{2}}
\end{equation}
where $E[\bullet]$ is the expected value operator, $j$ is the sample delay (or shift), $\mu$ and $\sigma$ are the mean and the standard deviation of $X$. Figure~\ref{fig:subfig1} shows the autocorrelation results of our raw-data. The low values of the autocorrelation between raw samples (8-bit per sample) verify the iid assumption of our physical model for min-entropy evaluation (see assumption 3 in Section~\ref{minentropy}). A slightly large coefficient at the 2nd delay sample is attributed to the finite bandwidth of our photodetector. After post-processing, the autocorrelation is substantially reduced as shown in Fig.~\ref{fig:subfig2}.

Some test results of the extracted data are given in Fig.~\ref{fig:subfig3} and Fig.~\ref{fig:subfig4}. With the sampling rate of 1 GHz, the corresponding random bit generation rate is over 6 Gb/s. We finally remark that our implementations of randomness extractors with Matlab on a standard PC are not fast enough for a real-time QRNG. In practice, this might restrict the random bit generation speed. It will be interesting for future investigations to create a real-time extractor (by a better software or hardware implementation) for our high-speed QRNG.

\section{Conclusion}
In conclusion, we have successfully demonstrated a high-speed QRNG
at a generation rate of over 6 Gb/s. The randomness is generated
from the intrinsic quantum phase fluctuations of spontaneous
emission photons. Our work not only highlights the importance on the
rigorous quantification and distillation of the quantum randomness
in a practical QRNG, but also demonstrates the large potential for
random number generations by quantum phase fluctuations as the true
entropy source. \\ \\
\textbf{Acknowledgments}
\\We thank enlightening discussions with V.~Burenkov, M.~Curty, Z.~Liao, C.~Rockoff, N.~Raghu, X.~Shan, C.~Weedbrook, J.~Xuan,
particularly L.~Qian and Z.~Yuan. H.~Xu and H.~Zheng are financially
supported by NSERC, USRA and CQIQC prized summer research
scholarship. Support from funding agencies NSERC, the CRC program,
CIFAR, and QuantumWorks is gratefully acknowledged.
\\ \\
\emph{Note added}: After posting of an early version of our paper on arXiv
\cite{Xu:QRNG}, we noticed that high-speed QRNGs based on different
approaches have recently appeared \cite{symul2011real, Jofre2011}.


\begin{thebibliography}{99}
\bibitem{temperature:controller}
The measured accuracy of the temperature controller is 0.01$^{\circ}C$. The fluctuations of the setpoint temperature of the PLC-MZI are
smaller than 0.01$^{\circ}C$ during a few hours.
\bibitem{experiment:gamma}
To experimentally determine $\gamma$, the key idea is that when the laser is operated at a significant high power level, the classical noise part ($C$ in Eq.~(\ref{phasefluctuations})) will dominate over the quantum fluctuations part ($\frac{Q}{P}$ in Eq.~(\ref{phasefluctuations})). It consists of three steps: a) at an optical power level $P_{o}$, we measured the variance of $V_{pr}(t)$ as $\sigma^{2}_{1}$. b) the laser was operated to its maximal power (around 25 mW for our DFB laser diode) and an optical attenuator (JDS Uniphase HA1) was applied right after the laser to attenuate the output power down to $P_{o}$, in which the variance of $V_{pr}(t)$ was measured as $\sigma^{2}_{2}$.  From $\sigma^{2}_{1}$ and $\sigma^{2}_{2}$, we could derive the experimental value $\gamma=\frac{\sigma^{2}_{1}-\sigma^{2}_{2}}{\sigma^{2}_{2}}$ at power $P_{o}$. c) the process was repeated at different power levels and the experimental results were shown in Fig.~\ref{ratio}.
\bibitem{adc:range}
We remark that in a practical system, it will be interesting for future research to investigate how to determine an optimal ADC range, which can maximize the extractable randomness.
\end{thebibliography}
\end{document}